\documentstyle[prc,preprint,aps]{revtex}
\begin{document}
\title{RPA approach to rotational symmetry restoration \\ 
in a three-level Lipkin model}
\author{K. Hagino and G.F. Bertsch}
\address{Institute for Nuclear Theory, Department of Physics, 
University of Washington, Seattle, WA 98195, USA}
\maketitle

\bigskip

\begin{abstract}

We study an extended Lipkin-Meshkov-Glick model that permits a 
transition to a deformed phase with a broken continuous symmetry.
Unlike simpler models, one sees a persistent zero-frequency
Goldstone mode past the transition point into the deformed phase.
We found that the RPA formula for the correlation energy provides 
a useful 
correction to the Hartree-Fock energy when the number of particle 
$N$ satisfies $N > 3$, and becomes accurate for large $N$. 
We conclude that the RPA correlation energy formula offers a
promising way to improve the Hartree-Fock energy in a systematic
theory of nuclear binding energies.

\end{abstract}   


\section{Introduction}

Hartree-Fock (HF) theory is the fundamental starting point
to understand the ground state properties of many-fermion systems. 
Its main assumption is that a particle independently moves in a 
mean field generated by other surrounding particles. 
In nuclear physics, with the adjustment of a few parameters of an 
effective interaction, the HF theory has described reasonably well the 
global properties of nuclei throughout the periodic table \cite{N82}.
However, correlation effects which go beyond HF are also significant.
They appear most dramatically when the HF ground state violates a
symmetry of the Hamiltonian such as rotational invariance or 
number conservation.  A global theory of nuclear binding must surely
take these correlation into account, if it is to achieve an accuracy
at the 1 MeV level.  A recipe is often 
used which is based on the projection after variation method \cite{RS80}. 

In this paper, we argue that the RPA correlation formula can 
provide a better method in correcting the broken symmetry in the HF 
theory. 
In the RPA, a restoration of the symmetry breaking appears as a 
zero energy solution of the RPA equations. The correlation energy 
associated with the several symmetries can be calculated using 
the RPA formula \cite{RS80,R68}
\begin{equation}
E_{corr}= \frac{1}{2}\left(\sum_i\hbar\omega_i - Tr(A)\right), 
\label{corr}
\end{equation}
where $\omega_i$ is the (positive) 
frequency of the RPA phonon for the $i$-th mode 
and $A$ is the $A$ matrix in the RPA equations. 
Our aim in this paper is to construct a simple Hamiltonian model to 
study the effects of correlations on the ground state energy and to 
show that the RPA formula is adequate in principle. 
To this end, we employ a three-level 
version \cite{LKD70,HY74,MKZ88,BPP95} of the 
Lipkin-Meshkov-Glick (LMG) model. 
The model describes $N$ identical fermions in three single-particle 
levels, each of which is $N$-fold degenerate. Exact solutions can be obtained 
by explicitly diagonalising the model Hamiltonian. The RPA correlation
energy was discussed by Brito {\it et al.}\cite{BPP95}, 
but their parameters did
not leave a continuous symmetry to be broken, as is the case for the
transition between spherical and deformed nuclei.
In this paper, we shall use parameters which mimic quadrupole motions 
in nuclei. In such a way, the correlation associated 
with rotational motion is easily studied. 

The paper is organised as follows. In Sec. II, we first show that 
the RPA formula is exact for a simple model 
with a two-body interaction, Eq. (\ref{Hho}) below. 
We then introduce the three-level Lipkin model in Sec. III and solve it 
in the HF as well as in the RPA. We compare the RPA correlation energy with 
the exact solution of the model and show that the RPA formula works well 
even in the vicinity of the critical point of the phase transition. 
Summary of the paper is given in Sec. IV, together 
with further discussions on the RPA formula. 

\section{RPA Correlation formula}

Before we study the correlation in the three-level Lipkin model, we 
would like to demonstrate that the RPA correlation formula 
works well, using an 
analytically solvable model. 
Consider a two fermion system bound in a 
harmonic potential coupled by a linear interaction,
\begin{equation}
H=H_0 + V = \sum_{i=1,2}\left(-\frac{\hbar^2}{2m}
\frac{\partial^2}{\partial x_i^2} + \frac{1}{2}m\omega_0^2 x_i^2\right)
-Cx_1x_2.
\label{Hho}
\end{equation}
This model was first 
introduced in Ref. \cite{EB83} to discuss the effects of the RPA 
correlation on the ground state density. A similar model has been 
considered in Ref. \cite{TEE98} in connection with the paired 
Wigner crystal. Using the transformation $\xi=(x_1 + x_2)/\sqrt{2}$ and 
$\eta=(x_1 - x_2)/\sqrt{2}$, 
the Hamiltonian can be written in the form 
\begin{equation}
H = 
-\frac{\hbar^2}{2m}\frac{\partial^2}{\partial \xi^2} 
+ \frac{1}{2}m\left(\omega_0^2 -\frac{C}{m}\right)\xi^2
-\frac{\hbar^2}{2m}\frac{\partial^2}{\partial \eta^2} 
+ \frac{1}{2}m\left(\omega_0^2 +\frac{C}{m}\right)\eta^2,
\end{equation}
from which we obtain the exact ground state energy  
\begin{equation}
E_{exact}=\frac{1}{2}\hbar\omega_0\left(\sqrt{1-\frac{C}{m\omega_0^2}}
+ \sqrt{1+\frac{C}{m\omega_0^2}}\right).
\label{Eexact0}
\end{equation}
The limit where the coupling constant $C$ is equal to $m\omega_0^2$ 
corresponds to a translationally invariant Hamiltonian with an
interaction ${1\over 2} m\omega_0^2 (x_1-x_2)^2$, giving
\begin{equation}
E_{exact} = \hbar \omega_0/\sqrt{2}.
\label{Eexact}
\end{equation}

Let us now solve the problem in the mean field approximation and then 
consider the RPA correlation energy. If the coupling constant $C$ 
is small, one can regard $V$ in the Hamiltonian (\ref{Hho}) as a 
residual interaction. 
The mean field Hamiltonian $H_0$ then has already been decoupled and 
we immediately obtain 
\begin{equation}
E_{MF}=\hbar\omega_0/2 + \hbar\omega_0/2 = \hbar\omega_0.
\label{EMF}
\end{equation} 
We define the RPA excitation operator as 
\begin{equation}
Q^{\dagger}=\sum_{i=1,2}\left(X_ia_i^{\dagger} - Y_i a_i\right),
\end{equation}
where $a^{\dagger}$ and $a$ are the creation and the annihilation 
operators of the unperturbative phonon, respectively. 
The RPA equation then reads
\begin{equation}
\left(
\begin{array}{cccc}
\hbar\omega_0 & -C\alpha_0^2 & 0 & -C\alpha_0^2 \\
-C\alpha_0^2 & \hbar\omega_0 & -C\alpha_0^2 & 0 \\
0 & C\alpha_0^2 & -\hbar\omega_0 & C\alpha_0^2 \\
C\alpha_0^2 & 0 & C\alpha_0^2 & -\hbar\omega_0 
\end{array}\right)
\left(\begin{array}{c}
X_1\\X_2\\Y_1\\Y_2
\end{array}\right)
=\hbar\omega
\left(\begin{array}{c}
X_1\\X_2\\Y_1\\Y_2
\end{array}\right),
\end{equation}
$\alpha_0$ being the amplitude of the zero point motion defined as 
$\sqrt{\hbar / 2m\omega_0}$. 
The solutions of this equation are found to be 
\begin{equation}
\hbar\omega=\pm\left\{\hbar\omega_0\sqrt{1+\frac{C}{m\omega_0^2}}, 
~~~\hbar\omega_0\sqrt{1-\frac{C}{m\omega_0^2}}\right\}.
\label{RPA2}
\end{equation}
We thus obtain 
\begin{equation}
E_{corr} = 
\frac{1}{2}\hbar\omega_0\left(\sqrt{1-\frac{C}{m\omega_0^2}}
+ \sqrt{1+\frac{C}{m\omega_0^2}}\right)-\hbar\omega_0, 
\end{equation}
which is precisely the needed correction to get the ground state energy
Eq. (\ref{Eexact0}) starting from the mean field 
energy Eq. (\ref{EMF}).  It reproduces Eq. (\ref{Eexact})
in the translationally invariant case.

It is interesting to compare the RPA approach with other ways of
dealing with correlation energies associated with broken symmetries.
In the case of center-of-mass motion, a recipe is often 
to subtract the expectation value of the center of mass 
operator from the mean field energy. 
With our Hamiltonian, this prescription gives
\begin{equation}
E_{cm}=-<MF|\frac{1}{2}\frac{(p_1 + p_2)^2}{2m}|MF>
=-\frac{1}{4}\hbar\omega_0.
\end{equation}
The total $E_{MF} + E_{cm} = 3\hbar\omega_0 / 4 $ is not exact, although
it is close to Eq. (\ref{Eexact}).
This study clearly shows that the RPA formula provides a much better 
method to calculate correlation energies. 

\section{Three-level Lipkin model}

We now consider RPA correlations in a three-level Lipkin model. 
Labelling the levels 0,1, and 2, 
we choose the Hamiltonian to be invariant
under transformations between 1 and 2.  The Hamiltonian we consider
can be expressed  

\begin{equation}
H=\epsilon(\hat{n}_1+\hat{n}_2)
-\frac{V}{2}(K_{1}K_{1}+K_{2}K_{2}+
K_{1}^\dagger K_{1}^\dagger+K_{2}^\dagger K_{2}^\dagger),
\label{Hlipkin}
\end{equation}
where
\begin{eqnarray}
\hat{n}_\alpha&=&\sum^N_{i=1}c_{\alpha i}^{\dagger}c_{\alpha i},
~~~~~\alpha = 0,1,2, \\
K_{\alpha}&=&\sum^N_{i=1}c_{\alpha i}^{\dagger}c_{0i}, 
~~~~~\alpha = 1,2. 
\end{eqnarray}

\subsection{Exact Solutions}

Since the Hamiltonian given by Eq. (\ref{Hlipkin}) couples symmetric 
states with respect to interchange of particles 
only with other symmetric states, a suitable basis for the exact 
diagonalisation of the Hamiltonian $H$ is given by \cite{HY74}
\begin{equation}
|n_1n_2> = \sqrt{\frac{(N-n_1-n_2)!}{N!n_1!n_2!}}(K_{1})^{n_1}
(K_{2})^{n_2}|00>.
\end{equation}
This is a simultaneous eigenstate of the number operators 
$\hat{n}_1$ and $\hat{n}_2$ with the eigenvalue of $n_1$ and $n_2$, 
respectively. 
The effect of the $K$ operators on the states is given by relations
such as
\begin{equation}
K_{1}|n_1n_2>=\sqrt{(N-n_1-n_2)(n_1+1)}|n_1+1,n_2>
\end{equation}
The matrix elements of $H$ can easily be calculated and are given by
\begin{eqnarray}
<n_1'n_2'|H|n_1n_2>&=&\epsilon(n_1+n_2)
\delta_{n_1',n_1}\delta_{n_2',n_2}  \nonumber \\
&&-\frac{V}{2}\left(\sqrt{(n_1+1)(n_1+2)(N-n_1-n_2)(N-n_1-n_2-1)}
\delta_{n_1',n_1+2}\delta_{n_2',n_2}\right.  \nonumber \\
&&~~~~~ +\sqrt{n_1(n_1-1)(N-n_1-n_2+1)(N-n_1-n_2+2)}
\delta_{n_1',n_1-2}\delta_{n_2',n_2} \nonumber \\
&&~~~~~ +\sqrt{(n_2+1)(n_2+2)(N-n_1-n_2)(N-n_1-n_2-1)}
\delta_{n_1',n_1}\delta_{n_2',n_2+2}  \nonumber \\
&&~~~~~\left. +\sqrt{n_2(n_2-1)(N-n_1-n_2+1)(N-n_1-n_2+2)}
\delta_{n_1',n_1}\delta_{n_2',n_2-2}\right). 
\end{eqnarray}
The dimension of the matrix to be diagonalised is $(N+1)(N+2)/2$. 
Further reduction can be achieved by considering that the Hamiltonian 
conserves the parity of each levels\cite{MKZ88}. 

\subsection{Hartree-Fock Approximation}

Let us now solve the problem in the Hartree-Fock approximation. We
consider a transformation of basis defined by operators $a_{\alpha i}$,
with $a_{0 i}$ representing the occupied orbital.  The HF state has the
form
\begin{equation}
|HF>=\prod^{N}_{i=1}a_{0i}^{\dagger}|>.
\end{equation}
and the transformation of basis is such as to minimize the expectation
of the Hamiltonian.  Let us write the transformation as 
\begin{equation}
\left(\begin{array}{c}
a_{0i}^{\dagger}\\a_{1i}^{\dagger}\\a_{2i}^{\dagger}
\end{array}\right)
=
\left(
\begin{array}{ccc}
\cos\alpha & \cos\beta\sin\alpha & \sin\beta\sin\alpha \\
-\sin\alpha & \cos\beta\cos\alpha & 
\sin\beta\cos\alpha \\
0& -\sin\beta & 
\cos\beta
\end{array}\right)
\left(\begin{array}{c}
c_{0i}^{\dagger}\\c_{1i}^{\dagger}\\c_{2i}^{\dagger}
\end{array}\right).
\label{HFsp}
\end{equation}
Using these relations, it is straightforward to evaluate the energy surface 
$E(\alpha,\beta)$ \\ = $<HF|H|HF>$ as 
\begin{eqnarray}
E(\alpha,\beta)&=&N\epsilon\sin^2\alpha-VN(N-1)\sin^2\alpha\cos^2\alpha.
\label{LipkinMF}
\end{eqnarray}

Note that the potential surface $E(\alpha,\beta)$ is independent of 
$\beta$ and thus totally flat in the $\beta$ direction for the 
rotationally invariant Hamiltonian. For simplicity, we particularly 
choose $\beta=0$ in constructing the HF single particle operators, 
Eq. (\ref{HFsp}). The HF Hamiltonian thus spontaneously breaks the rotational 
symmetry, and the Goldstone mode will appear at zero excitation energy 
to restore the symmetry breaking, as we will show in the next subsection. 

The optimum choice of $\alpha$ is obtained by minimising the potential 
surface $E(\alpha,\beta)$.  It is convenient to express the solution in
term of the dimensionless parameter
\begin{equation}
\chi\equiv V(N-1)/\epsilon.
\end{equation}
For $\chi < 1$, the minimum appears at 
$\alpha=0$ (spherical phase). At $\chi=1$, the system undergoes a phase 
transition and, for $\chi > 1$, the potential surface displays two symmetrical 
minima at $\cos2\alpha=1/\chi$ (deformed phase).
The ground state energy in the HF approximation is thus given by
\begin{equation}
E_{HF}=\left\{
\begin{array}{ll}
0&\qquad (\chi < 1), \\
\noalign{\vskip 0.2 cm}
\displaystyle{\frac{N\epsilon}{4}\left(2-\chi-1/\chi\right)} 
&\qquad (\chi > 1).
\end{array}\right.
\label{EMF2}
\end{equation}

\subsection{Random Phase Approximation}

We next solve the problem in the RPA in order to evaluate the correlation 
energy associated with the rotational motion. We define the RPA excitation 
operator as
\begin{equation}
Q^{\dagger}=X_1\widetilde{K}_{1}+X_2\widetilde{K}_{2}
-Y_1\widetilde{K}_{1}^\dagger-Y_2\widetilde{K}_{2}^\dagger,
\end{equation}
where 
\begin{equation}
\widetilde{K}_{\alpha}=\sum^N_{i=1}a_{\alpha i}^{\dagger}a_{0i}, 
~~~~\alpha = 1,2.
\end{equation}
The RPA equation is obtained from $<HF|[\delta Q,[H,Q^{\dagger}]-\omega
Q^{\dagger}]|HF>=0$ for $\delta Q=\widetilde{K}_{1},\widetilde{K}_{2},
\widetilde{K}_{1}^\dagger$, and $\widetilde{K}_{2}^\dagger$. The result
is the well-known RPA matrix equation
\begin{equation}
\left(
\begin{array}{cc}
A & B \\
-A & -B 
\end{array}\right)
\left(\begin{array}{c}
X \\ Y
\end{array}\right)
=
 \omega \left(\begin{array}{c}
X  \\ Y
\end{array}\right),
\end{equation}
where $A$ and $B$ are 2$\times$2 matrices given by
\begin{eqnarray}
A_{11}&=&\epsilon\cos2\alpha+\frac{3}{2}\epsilon\chi\sin^22\alpha, \\
A_{12}&=&A_{21}=0, \\
A_{22}&=&\epsilon(1-\sin^2\alpha)+\frac{1}{2}\epsilon\chi\sin^22\alpha,\\
B_{11}&=&-\epsilon\chi(\cos^4\alpha+\sin^4\alpha),\\
B_{12}&=&B_{21}=0, \\
B_{22}&=&-\epsilon\chi\cos^2\alpha.
\end{eqnarray}
Because $A$ and $B$ are separately diagonal, the RPA matrix can be easily
diagonalized. The RPA frequencies are found to be
\begin{eqnarray}
\omega_1^2&=&(A_{11}+B_{11})(A_{11}-B_{11}), \\
\omega_2^2&=&(A_{22}+B_{22})(A_{22}-B_{22}). 
\end{eqnarray}
Substituting the self-consistent value for $\alpha$ 
obtained in the previous sub-section, we obtain 
\begin{eqnarray}
\omega_1^2&=&
\left\{\begin{array}{ll}
\epsilon^2(1+\chi)(1-\chi)& \qquad (\chi < 1), \\
\noalign{\vskip 0.2 cm}
2\epsilon^2(\chi+1)(\chi-1)&\qquad (\chi > 1),
\end{array}\right. \\
\omega_2^2&=&
\left\{\begin{array}{ll}
\epsilon^2(1+\chi)(1-\chi)& \qquad (\chi < 1), \\
\noalign{\vskip 0.2 cm}
0 &\qquad (\chi > 1).
\end{array}\right. 
\end{eqnarray}
In the spherical phase, the RPA frequencies for the two modes 
are identical. In the deformed phase, on the other hand, the frequency for the 
second mode becomes zero. In this case, the first mode corresponds to the 
beta vibration, while the second mode corresponds to the 
rotational motion perpendicular to the symmetry axis. 
Figure 1 shows the RPA frequencies as a function of $\chi$ for $N=20$. 
One can clearly see the discontinuity at the critical point $\chi=1$.

Figure 2 compares the ground state energy as a function of $\chi$ 
obtained by several methods. The number of particle $N$ is set to be 20. 
The solid line is the exact solution obtained by numerically 
diagonalising the Hamiltonian. The dashed line is the ground state energy 
in the Hartree-Fock approximation given by Eq. 
(\ref{EMF2}). It considerably deviates from the exact solution through 
the entire range of $\chi$ shown in the figure. The dot-dashed line 
takes into account the RPA correlation energy in addition to the HF 
energy. The improvement is apparent and significant. It is remarkable that 
the RPA formula works well even in the vicinity of the critical point of the 
phase transition $\chi=1$. 

It is expected that the RPA should be accurate for a large number of 
particles, and $N=20$ seems to fulfil that condition. 
We also would like to apply the approximation to nuclei with just a few 
valence nucleons. Figure 3 compares the exact energy with the HF + RPA 
as a function of $N$, for $\chi$=5.0. The RPA correction may be considered 
useful if it is within a factor of 2 of the exact. We see from the figure 
that this is satisfied for $N\geq 4$. For $N=2$, the exact correlation 
energy is several times RPA, and such near-magic nuclei would require a more 
elaborate way. 

\section{Summary and Discussions}

We discussed the role played by the RPA correlation in the ground state 
energy. To this end, we used simple Hamiltonian models, like a bi-linear 
interaction between two fermions as well as a three-level Lipkin model. 
The former is for the correlation associated with the center of mass motion, 
while the latter for the rotational motion. 
We showed that the ground state energy is well described in the mean field 
theory once the ground state correlations are taken into account in the 
RPA. We also showed that the RPA formula works well for a wide region of 
an order parameter, including in the vicinity of the critical point of the 
phase transition. 
Evidently, 
the RPA formula provides a powerful 
method to calculate energies for long range correlations, which are not 
included in the HF approximation. 

Up to now, microscopic theory based on the mean field theory 
has not been as successful than other approaches 
in making a global fit to nuclear binding energies. The most accurate
theory of nuclear binding systematics \cite{mo95} starts from the 
liquid drop model, and treats shell effects perturbatively.  It fits
the binding energies with an RMS deviation of 
0.67 MeV, a factor of three better than the Gogny or the published Skyrme 
functional. 
We believe that there are good prospects to develop a better microscopic 
global theory, treating correlation energies systematically
by the RPA. A work towards this direction is now in progress \cite{BHR99}. 

\section*{Acknowledgement}
G.F.B. acknowledges support from the U.S.
Dept. of Energy under Grant DE-FG-06ER46561.


\bigskip

\begin{center}
{\bf Figure Captions}
\end{center}

\noindent
{\bf Fig.1:} 
RPA frequencies for collective vibrations as a function of 
$\chi\equiv V(N-1)/\epsilon$. The number of particle $N$ is 
chosen to be 20. 

\medskip

\noindent
{\bf Fig.2:} 
Comparison of the ground state energy obtained by several methods. 
The solid line is the exact numerical solution. The ground state energy 
in the Hartree-Fock approximation is denoted by the dashed line, while 
the dot-dashed line takes into account the RPA correlation energy in 
addition to that. 

\noindent
{\bf Fig.3:} 
The ground state energy as a function of $N$ for $\chi=5.0$. 
The meaning of each line is the same as in Fig. 2. 

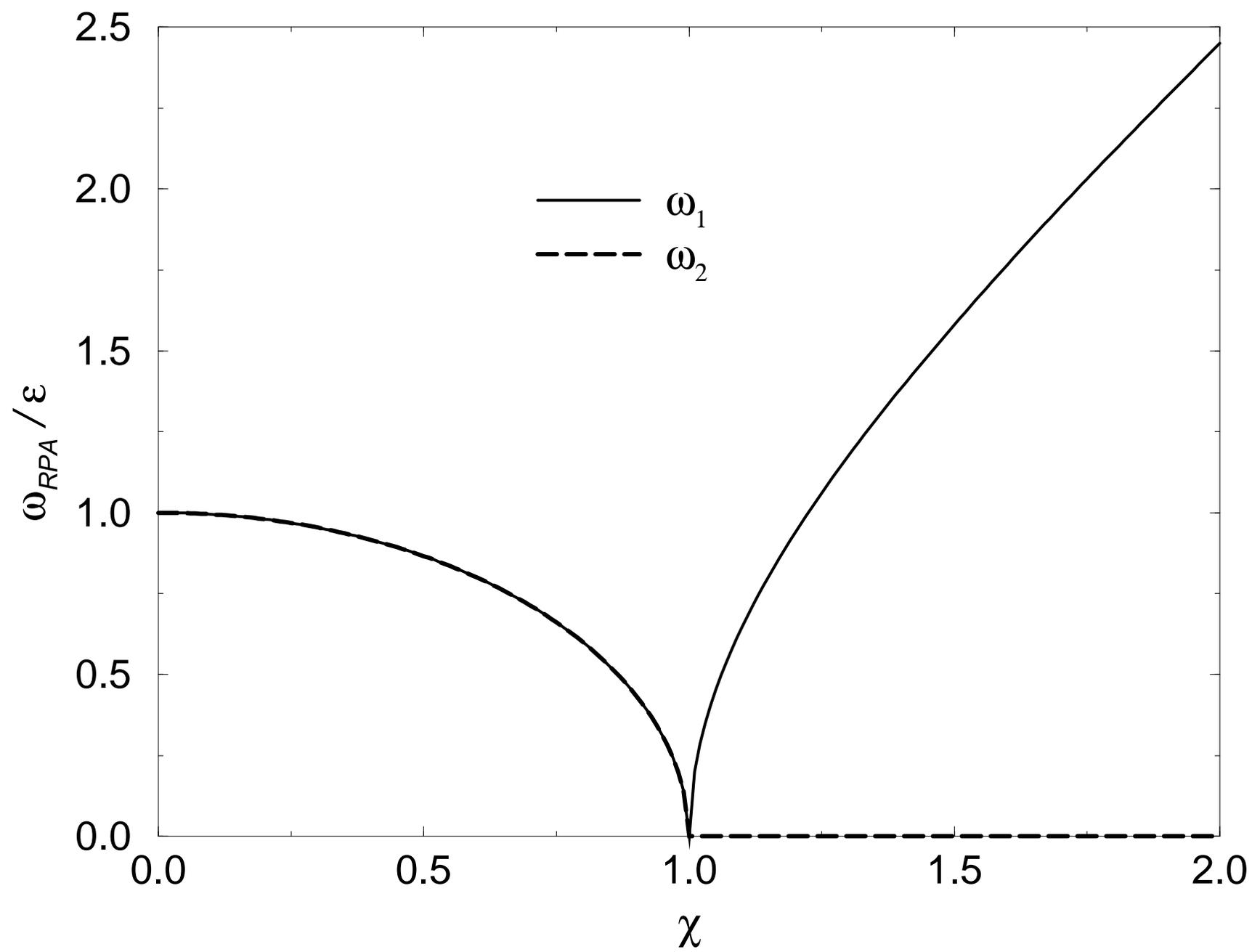
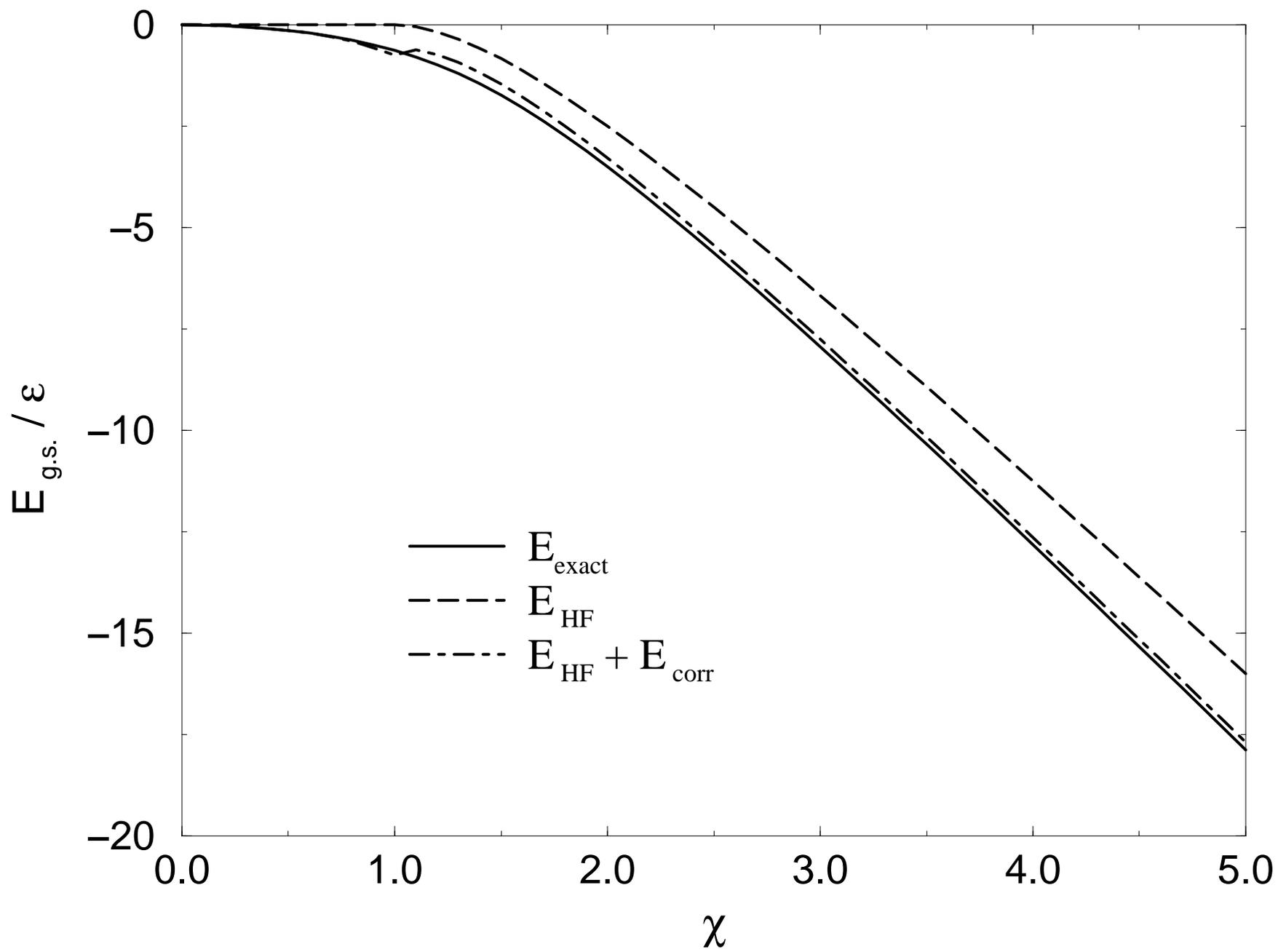
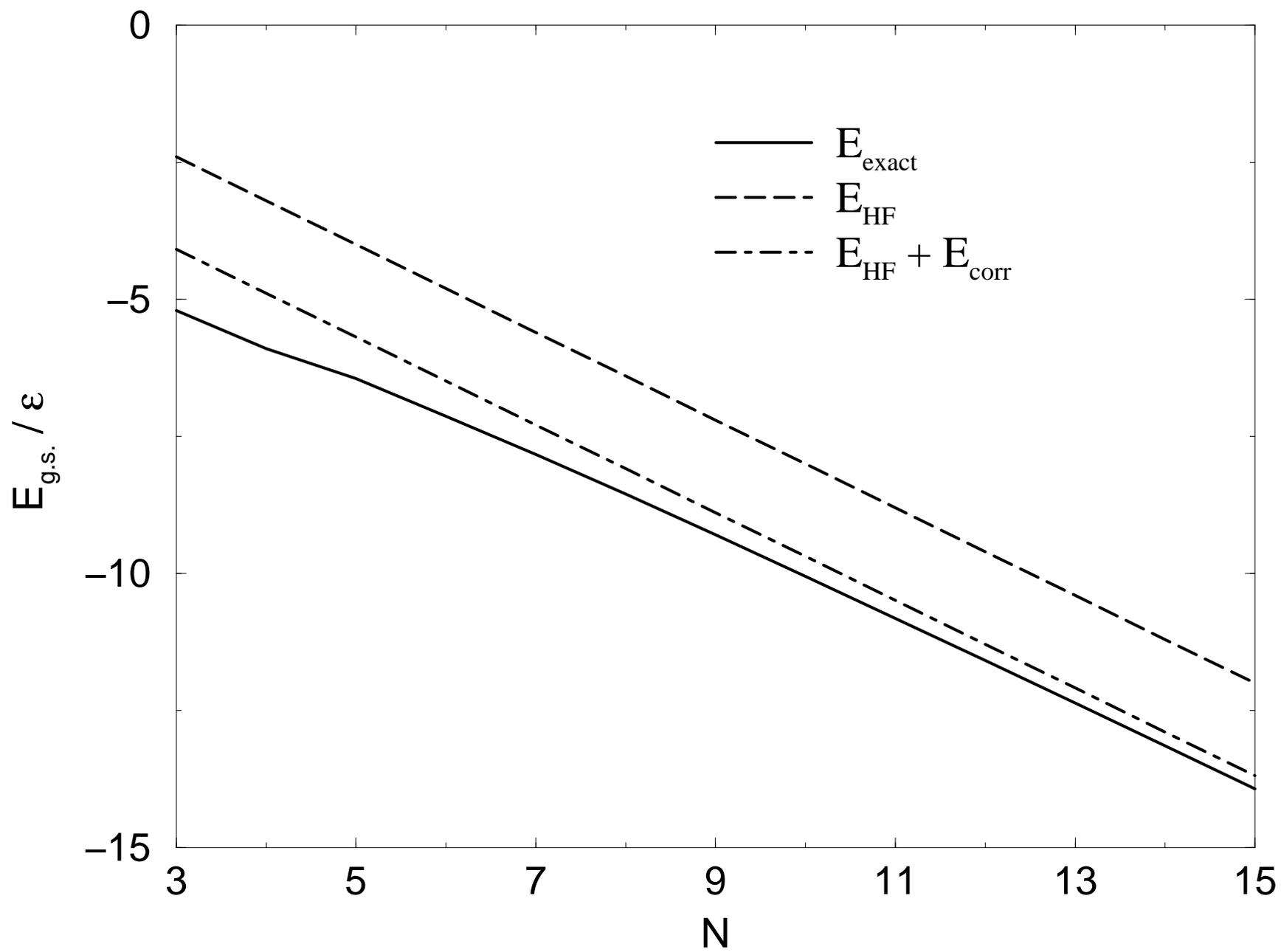
\end{document}